\documentclass[times]{kristall}     

\usepackage{epsf}
\usepackage[round]{natbib}          

\begin{document}
\maketitle

\begin{abstract}
\Abstract
\end{abstract}


\def\St{\large\strut}
\def\dc{C$^\circ$}
\def\tc{T$_{c}$}
\def\cuau{Cu$_{3}$Au\ }
\def\oh{$\frac{1}{2}$}
\def\th{$\frac{3}{2}$}
\def\u2{$\langle u^{2} \rangle$}
\renewcommand{\sb}[1]{$_{#1}$}

\newlength{\picwidth}
\setlength\picwidth {0.85\linewidth}


%
\section*{Introduction}
In many alloys the different atoms are not distributed randomly, but show
chemical short range order (SRO). The only information that can be
obtained from a routine crystallographic structure refinement based on
Bragg intensities is the average occupancies of the different atomic
sites. The chemical SRO shows up as diffuse scattering. This diffuse
scattering, obtained from single crystal scattering experiments, can be
analyzed and SRO parameters can be obtained. A summary of this method can
be found e.g. in \citet{cowley}. However, in some cases it might be
impossible to grow single crystals or to obtain crystals of sufficient
quality for single crystal diffuse scattering measurements. In these cases
a viable alternative is the use of powder diffraction as we demonstrate in
this paper. A recently emerging method to analyze the local structure
based on the total scattering pattern from crystalline powder samples is
the atomic pair distribution function (PDF) method. The PDF is obtained
via Fourier transform from the scattering data containing both Bragg and
diffuse scattering contributions. The PDF gives the probability to find an
atom at a distance r from another atom, in other words it gives the bond
length distribution of the material. This approach is long known in the
field of studying short range order in liquids and glasses but has
recently been applied to crystalline materials showing positional disorder
(e.g. \citet{egami;lsfd98,prdibi99,peprje00,bobikw99,loeg99,pebihe00}, for
more general information see \citet{www;total}). However, it has been
unclear if the PDF would be sensitive enough to extract information about
chemical ordering in perfectly crystalline materials. In a previous study
\citep{pr00} we have used simulated data based on structures showing
chemical SRO to test the ability of the PDF method to extract SRO
parameters. Those tests were successful and showed that in principle such
information can be extracted. In this paper we chose the system \cuau to
verify those findings using 'real' data. The chemical ordering in \cuau
has been studied for quite some time (e.g. \citet{chcoco70}). The
structure of \cuau is cubic. The system undergoes a chemical
order-disorder phase transition at \tc=394~\dc. Above \tc\ the two atom
types are randomly distributed, below \tc\ copper atoms occupy the corners
of the unit cell and gold occupies the face centers. This work should be
seen as proof of principle and we concentrate on the simplest case
comparing a quenched sample showing nearly random ordering of Cu and Au
with the fully ordered case.
%
%
\section*{Experiment}
\begin{figure}[!tb]
  \epsfxsize=\picwidth \epsfbox{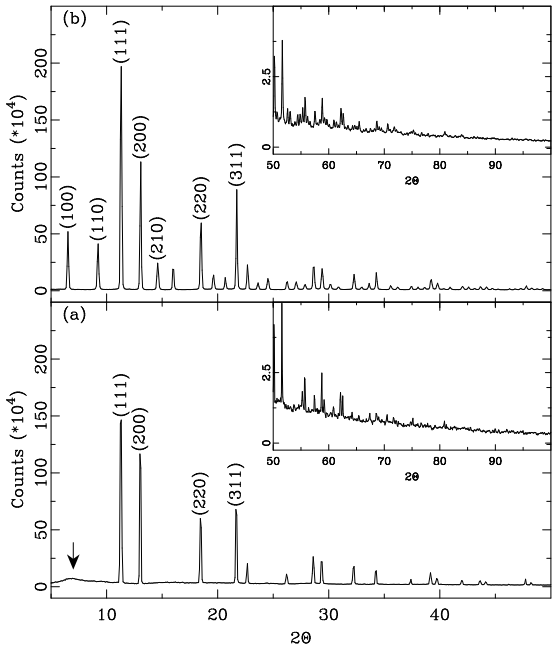}
  \caption[] {Experimental X-ray powder diffraction data for (a) ordered
              \cuau and (b) disordered \cuau. Peaks are labelled with the
              corresponding Miller indexes.}
  \label{fig;raw}
\end{figure}

Two samples were investigated: ordered and disordered \cuau. Both were
obtained from the same master alloy prepared by melting of 99.999 \% Cu
and Au under vacuum. The alloy was homogenized by annealing for 2~h at
1200~K. An ingot of the annealed master alloy was rapidly quenched down to
room temperature resulting in disordered \cuau. Another ingot was slowly
cooled down and allowed to anneal for a week at 723~K. This allowed the
material to reach an ordered state. Both samples were then filed into
powder. The powder samples were carefully packed between Kapton foils to
avoid texture formation and subjected to diffraction experiments using
x-rays of energy 29.09~keV ($\lambda$=0.4257\AA). The measurements were
carried out in symmetric transmission geometry at the beamline X7A of the
National Synchrotron Light Source (NSLS), Brookhaven National Laboratory.
Scattered radiation was collected with an intrinsic germanium detector
connected to a multichannel analyzer. Experimental powder diffraction
patterns are shown in Fig. \ref{fig;raw}. These data were then normalized
in electron units and converted to the structure functions $S(Q)$
\citep{klug} after due correction for flux, background, Compton
scattering, and absorption. The corresponding atomic pair distribution
(PDF) functions, G(r), are shown in Fig. \ref{fig;comp}. The data for both
samples were terminated at $Q_{\mbox{max}}=22.5$\AA$^{-1}$. All
corrections and data processing was done with the help of the program
\textit{RAD} \citep{pe89}. It is clear that, as well as the diffraction
patterns, the PDFs of the ordered and disordered samples are significantly
different.
%
%
\section*{Modelling and Results}
\begin{figure}[!tb]
  \epsfxsize=\picwidth \epsfbox{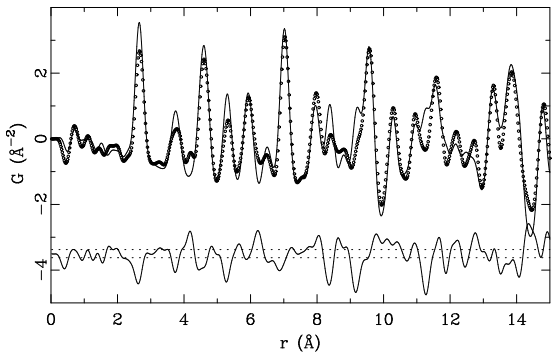}
  \caption[] {Experimental PDFs from disordered (circles) and ordered
              (solid line) \cuau. The difference is given below as solid
              line. The estimated experimental uncertainty is
              marked by horizontal broken lines.}
  \label{fig;comp}
\end{figure}
\begin{figure}[!tb]
  \epsfxsize=\picwidth \epsfbox{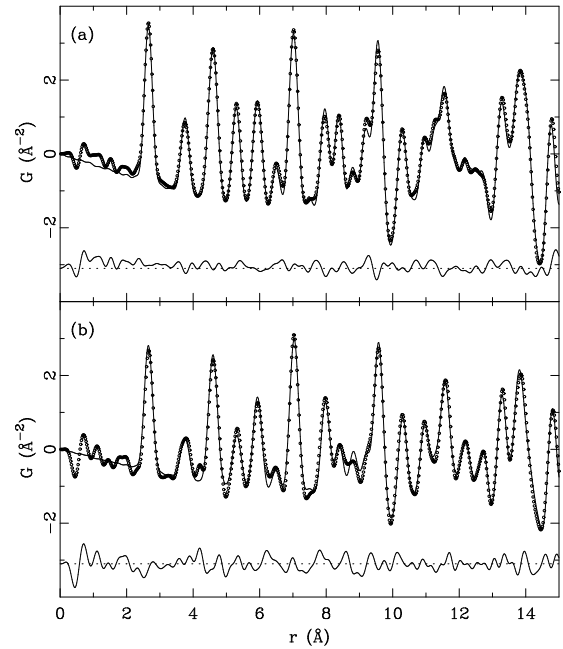}
  \caption[] {Experimental (circles) and calculated (solid line) PDFs for
              the ordered \cuau sample (a) and the disordered sample (b).
              The residual differences are shown below the data in both
              panels.}
  \label{fig;gr}
\end{figure}
\begin{table}[!tb]
  \caption[]{Structural data of ordered and disordered \cuau from
             PDF refinements. In the ordered case Cu is at (\oh,0,\oh) and
             Au is at (0,0,0). For the disordered sample Cu and Au occupy
             both sites with an occupancy of 0.75 and 0.25, respectively.
             The units for the lattice parameters are \AA\ and for values
             of \u2 \AA$^{2}$. The numbers in parentheses are the estimated
             standard deviation on the last digit. The weighted R-value
             R$_{wp}$ corresponds to the refinement of the average structure,
             R$_{wp}$ (RMC) corresponds to the result of the RMC refinement.
             For more details see text.} \small
  \label{tab;struc}
  \begin{tabular*}{\linewidth}{@{\extracolsep{\fill}} @{}ccc}
  \hline
                             &  \textbf{ordered sample}
                             &  \textbf{disordered sample}            \St \\
  \hline
  a                          &   3.75162(11)            &   3.7585(3) \St  \\
  \u2(Cu/Au)                 &   0.01307(2)/0.00772(3)  &   0.00939(4)\St  \\
  R$_{wp}$/R$_{wp}$(RMC)     &   9.5/9.5                &  16.2/15.4  \St  \\
  \hline
  \end{tabular*}
\end{table}
Bragg peaks in the diffraction pattern of disordered \cuau can be indexed
with a cubic unit cell, space group $Fm\overline{3}m$. The ordered sample
can be indexed in space group $Pm\overline{3}m$. The Miller indices (hkl)
are shown in Fig. \ref{fig;raw}. One observes additional reflections for
the ordered sample with mixed (odd + even) values of (hkl). These
reflections can be interpreted as super-lattice peaks showing up as a
result of the ordering of copper and gold. The degree of order in the
sample can be quantified by comparing the intensities of the super-lattice
and fundamental Bragg peaks \citep{suryanarayana,guinier}. However, in
cases where chemical SRO is confined to only a few nearest neighbors, one
observes only broad diffuse scattering rather than sharp super-lattice
peaks. A broad bump in the diffraction pattern at $2\Theta \approx 7^{o}$
marked by an arrow in Fig. \ref{fig;raw}a suggests the presence of SRO
diffuse scattering. We chose to use the PDF technique to extract SRO
parameters. The PDFs of the ordered and disordered sample are shown in
Fig. \ref{fig;comp}. The PDF peaks for both samples are in nearly the same
positions, since copper and gold are occupying the same cubic lattices.
However, the intensity of the PDF peaks are different, since the PDF peak
intensities are weighted by the respective scattering powers of the atoms
contributing to that interatomic distance. Fig. \ref{fig;comp} also shows
that the difference is significantly larger than the experimental
uncertainty shown as dashed lines.
\par

The modelling of the PDF data is carried out in two steps: First the PDFs
are refined based on a model of the average crystal structure. In a second
step Cu and Au are allowed to switch sites in a Reverse Monte Carlo (RMC)
refinement \citep{mcpu88} of a large model crystal. Chemical SRO
parameters are then extracted from the final configuration of the RMC
refinement.
\par

The refinements of the average structure were carried out using the
program \textit{PDFFIT} \citep{prbi99}. The refined PDFs are shown in Fig.
\ref{fig;gr}. The only structural parameters refined are the lattice
parameter and atomic displacement parameters for Cu and Au. These
parameters are listed in Tab. \ref{tab;struc}. Details about the
refinement procedure can be found in \citet{prbi99} and are not discussed
here. The refinement extended over a range of $2.1<r<15.0$~\AA. It is
obvious from Fig. \ref{fig;gr} as well as the weighted R values listed in
Tab. \ref{tab;struc} that the refinement corresponding to the random
sample is not as good as for the ordered sample. When trying to refine
site occupancies for the random sample, their values stayed at the
expected 0.75/0.25 for Cu and Au respectively within one standard
deviation. However, the worse R value as well as the modulated background
of the scattering data for the random sample (Fig. \ref{fig;raw}), suggest
some degree of chemical SRO present in the quenched sample, referred to as
random in this paper. For the ordered sample, we observe no such
indication.
\par

The next step of the analysis are RMC refinements to allow Cu and Au to
order locally. A comprehensive summary of the method applied to
crystalline materials is given in \citet{tudoke01}. The algorithm works as
follows: First, the PDF is calculated from the chosen crystal starting
configuration and a goodness-of-fit parameter $\chi^{2}$ is computed.
\begin{equation}
    \chi^{2} = \sum_{i=1}^{N}
               \frac{ ( G_{e}(r_{i}) - G_{c}(r_{i})) ^{2}}
                    {\sigma^{2}}
    \label{eq;rmc}
\end{equation}
The sum is over all measured data points $r_{i}$, $G_{e}$ stands for the
experimental and $G_{c}$ for the calculated PDF.  The RMC simulation
proceeds with the selection of a random site within the crystal.  The
system variables associated with this site, such as occupancy or
displacement, are changed by a random amount, and then the model PDF and
the goodness-of-fit parameter $\chi^{2}$ are recalculated.  The change
$\Delta\chi^{2}$ of the goodness-of-fit $\chi^{2}$ before and after the
generated move is computed. Every move which improves the fit
($\Delta\chi^{2} < 0$) is accepted.  'Bad' moves worsening the agreement
between the observed and calculated PDF are accepted with a probability of
$P=\exp(-\Delta\chi^{2}/2)$. For more details about the RMC refinement
procedure used, the reader should refer to \citet{pr00} and the references
therein. All RMC refinements were carried out using the program
\textit{DISCUS} \citep{prne97}. We used a model crystal of 20x20x20 unit
cells in size containing a total of 32000 atoms. The starting structure
was given by the refinement result of the average structure (see Tab.
\ref{tab;struc}). The only allowed RMC move was to have Cu and Au change
sites. This way the overall concentration of Cu and Au was preserved. The
refinement was aborted after 15 cycles when no further improvement in
$\chi^2$ was observed. A cycle is defined as the number of RMC moves
required to visit every atom in the model crystal once on average.
\par
To analyze the resulting structure after the RMC refinement is most
convenient to describe the chemical SRO using correlation coefficients
$c_{ij}$ which are defined as:
\begin{equation}
  c_{ij} = \frac {P_{ij} - \theta^{2}} { \theta (1 - \theta)}
  \label{eq;corr}
\end{equation}
$P_{ij}$ is the joint probability that both sites $i$ and $j$ are
occupied by the same atom type and $\theta$ is its overall occupancy.
Negative values of $c_{ij}$ correspond to situations where the two
sites $i$ and $j$ tend to be occupied by {\it different} atom types
while positive values indicate that sites $i$ and $j$ tend to be
occupied by the {\it same} atom type.  A correlation value of zero
describes a random distribution.  The maximum negative value of
$c_{ij}$ for a given concentration $\theta$ is $-\theta/(1-\theta)$
($P_{ij}=0$), the maximum positive value is +1 ($P_{ij}=\theta$).
\par

Table \ref{tab;corr} lists the correlation parameters for the first eight
neighboring shells for the ordered and disordered sample. The columns
labelled 'expected' give the theoretical correlation parameters for the
fully ordered or fully random case. The columns labelled 'RMC' list the
resulting correlation coefficients of the RMC refinements. In case of the
ordered sample, the RMC refinement was not able to find any configuration
different from the fully ordered case that would lead to an improvement of
$\chi^2$. As a result, the resulting correlation parameters are identical
to the expected ones. In case of the random sample the situation is
different, we observe values significantly different from zero which
describes the fully random case. The weighted R value improved from 16.2\%
to 15.4\% in the RMC refinement of the PDF from the random sample.
Inspection of Tab. \ref{tab;corr} shows that the chemical SRO found in the
sample refereed to as random is in fact of the same nature as in the
ordered sample, i.e. nearest Cu-Cu and Au-Au neighbors are avoided and
second nearest neighbors are preferred. However, for neighboring shells
higher than five, the order is indeed random. In other words, locally Cu
and Au atoms started ordering below the transition temperature, \tc,
before being arrested by the rapid quench.
\begin{table}[!bt]
\caption[]{Cu$_{3}$Au RMC refinement results. For
           details see text.}\small
  \begin{tabular*}{\linewidth}{@{\extracolsep{\fill}} @{}crrrr}
  \hline
       &  \multicolumn{2}{c}{\textbf{ordered sample}}  &
          \multicolumn{2}{c}{\textbf{disordered sample}}\St\\
  Neighbor & expected & RMC & expected & RMC \St\\
  \hline
  \oh\oh 0& -0.33 & -0.33 &  0.00 & -0.10\St\\
  100     &  1.00 &  1.00 &  0.00 &  0.19\St\\
  1\oh\oh & -0.33 & -0.33 &  0.00 & -0.05\St\\
  110     &  1.00 &  1.00 &  0.00 &  0.13\St\\
  \th\oh 0& -0.33 & -0.33 &  0.00 & -0.10\St\\
  111     &  1.00 &  1.00 &  0.00 &  0.05\St\\
  \th 1\oh& -0.33 & -0.33 &  0.00 & -0.01\St\\
  200     &  1.00 &  1.00 &  0.00 & -0.08\St\\
  \hline
  \end{tabular*}
\label{tab;corr}
\end{table}
%
%
\section*{Conclusion}
\begin{figure}[!tb]
  \epsfxsize=\picwidth \epsfbox{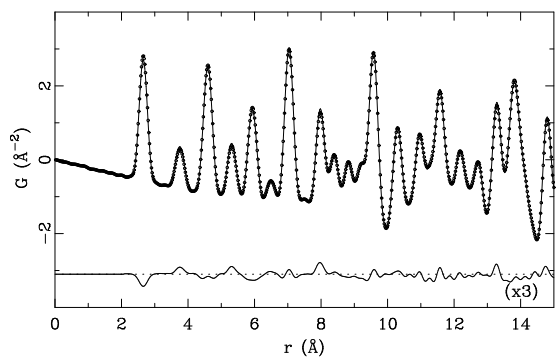}
  \caption[] {Calculated PDFs of the starting (solid line) and resulting
              structure (circles) of \cuau from the RMC refinement. The
              difference between both PDFs is enlarged by a factor of three
              and shown below the curves.}
  \label{fig;rmc}
\end{figure}
In an earlier paper \citep{pr00} we have shown that chemical SRO
parameters can successfully be extracted from simulated data using the PDF
analysis method. In this paper we have shown that chemical SRO parameters
can also successfully be extracted from real data. The results show that
in our rapidly quenched \cuau sample, locally there is a tendency for Cu
and Au to order in a way characteristic for the fully ordered sample.
\par
Two requirements can be identified to successfully extract SRO parameters
from PDF data: First high quality data need to be obtained. The
information about the SRO is only a small fraction of the PDF intensities,
obviously depending on the particular scattering powers of the sample
under investigation. Fig. \ref{fig;rmc} shows the difference between the
calculated PDF of the starting structure of the random sample and the
resulting PDF after the RMC refinement. The extracted information of the
Cu-Au ordering is contained in the difference of the two. Note that the
difference curve shown in Fig. \ref{fig;rmc} is magnified by a factor of
three. The other important factor is to have sensible constraints when
running the RMC refinement. This is discussed in much detail in the paper
by \citet{tudoke01}. In this case the RMC refinement was only allowed to
switch sites for copper and gold atoms. The structural parameters obtained
by the refinement of a model of the average structure were kept fix. This
way the only way to improve the fit was to introduce Cu-Au SRO. As a next
step, we are planning to collect in-situ high temperature PDFs of \cuau in
order to be able to compare the extracted SRO parameters with literature
values obtained from single crystal diffuse scattering data. It is clear
that there are also many ways how the data refinement can be improved. In
our example, the displacement parameters for the random sample were
identical for copper and gold, where as they are quite different in the
ordered case. One possible step forward would be to allow one to adjust
site specific displacement parameters during the RMC refinement. Efforts
in this direction are under way. In our previous paper \citep{pr00} we
have also investigated cases where chemical SRO as well as local
distortions are present in the sample. Our efforts now focus on finding a
suitable system to demonstrate that the PDF method also works in this more
complicated case.
\par
In summary, we have demonstrated that it is possible to extract chemical
SRO parameters of crystalline materials from the PDF obtained from powder
diffraction data. This opens the door for rapid determination of SRO
parameters as a function of external parameters such as temperature or
pressure using the PDF technique.
%
%
\begin{acknowledgment}
We are very grateful to Marin Gospodinov for making the \cuau samples.
This work was in part supported by DOE through Grant No. DE FG02
97ER45651. This work benefitted from the use of the National Synchrotron
Light Source (NSLS).  This facility is funded by the U.S. Department of
Energy under contract DE-AC02-98CH10886. Los Alamos National Laboratory is
funded by the US Department of Energy under contract W-7405-ENG-36.
\end{acknowledgment}
%
%
%


\end{document}